# Simultaneous Coherent and Displacement Compounding for 2D Noninvasive Carotid Strain Imaging: a Proof of Principle Study

Moein Mozaffarzadeh, Anne E.C.M. Saris, Jan Menssen, Chris L. de Korte

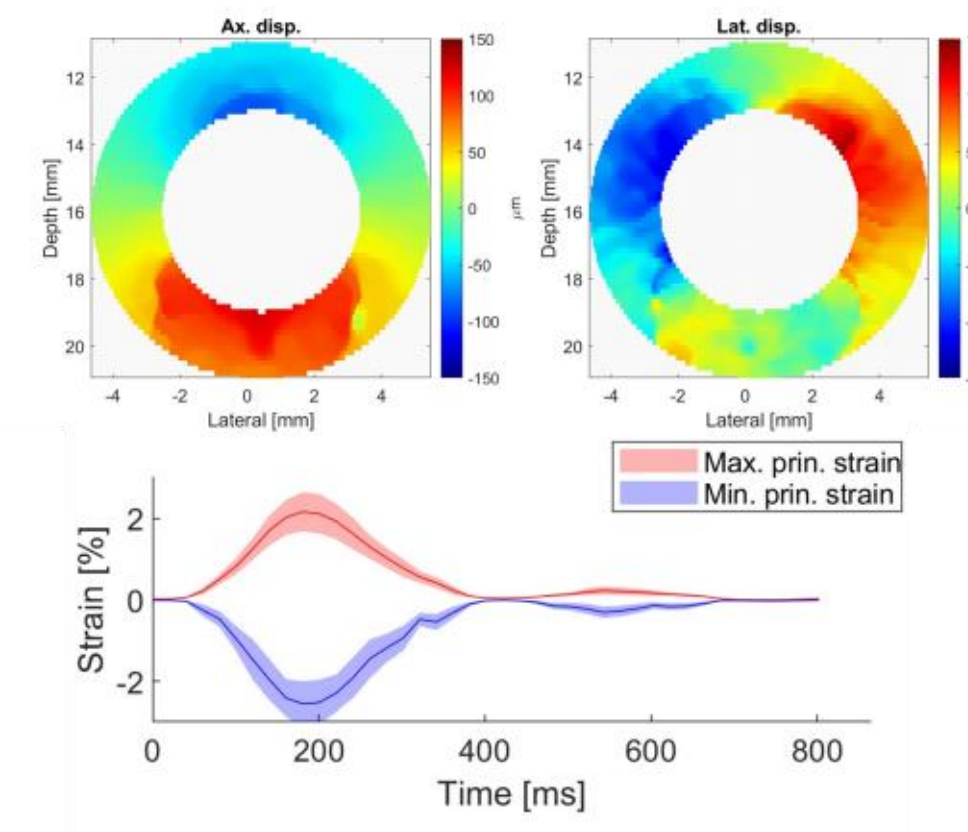

***Abstract*—Arteriosclerosis results from lipid buildup in artery walls, leading to plaque formation, and is a leading cause of death. Plaque rupture can cause blood clots that might lead to a stroke. Distinguishing plaque types is a challenge, but ultrasound elastography can help by assessing plaque composition based on strain values. Since the artery has a circular structure, an accurate axial and lateral displacement strategy is needed to derive the radial and circumferential strains. A high precision lateral displacement is challenging due to the lack of phase information in the lateral direction of the beamformed RF data. Previously, our group has developed a compounding technique in which the lateral displacement is estimated using tri-angulation of the axial displacement estimated from transmitting and beamforming ultrasound beams at ±20°. However, its applicability to *in vivo* is challenging due to the imaging noise and the low contrast of the arterial wall, caused by a single plane wave transmission. In this paper, we combine our displacement compounding with coherent compounding. Instead of transmitting a single plane wave, multiple plane waves are transmitted at certain angles with respect to the angle of the beamforming grids, and then the backscattered wavefronts are beamformed and coherently compounded on the center of the transmit beams (-20°, +20° and 0°). The numerical investigation using the GE9LD probe ($f_0$ = 5.32 MHz, pitch = 230 µm, width = 43.9 mm) led us to 19 plane waves spanning angles within -10° to 10° (with respect to center of the transmit beams); resulting in a total of 57 plane wave transmit (for 3 beamforming grids at 0° and ±20°). FIELD II simulations of a cylindrically-shaped phantom (mimicking the carotid artery) at a signal-to-noise ratio ≥ 20 dB shows that the proposed method decreases the root-mean-square-error of the lateral displacement and strain estimations by 40% and 45% compared to the previous method, respectively. The results of our experiments with a carotid artery phantom (made out of 10% PVA) show that the proposed method provides strain images with a higher quality and more in agreement with the theory.**

***Index Terms*—Strain imaging, coherent compounding, displacement estimation, plane wave transmission, carotid artery.**

## I. Introduction

ARTERIOSCLEROSIS is a leading cause of death. It initially results from the accumulation of excess fatty droplets in the intima of vessel walls. The growth of fatty droplets gets promoted by inflammatory molecules and results in a lipid pool, that is covered by a fibrous cap, resulting in a so-called atherosclerotic plaque. These plaques are considered to be rupture prone in contrary to plaques that are predominantly fibrous and calcified [1, 2]. As the intraluminal pressure fluctuates throughout the cardiac cycle, it exerts force on the artery wall. It is postulated that these forces can surpass the maximum strength of the thin fibrous cap, leading to the rupture of plaques [3, 4]. When a rupture occurs, the thrombogenic plaque material is exposed to the bloodstream leading to the formation of a blood clot (thrombus). Consequently, this thrombus can travel through the bloodstream and cause an acute occlusion of downstream cerebral branches, resulting in either a stroke or a transient ischemic attack [5, 6]. When plaque rupture occurs in one of the main coronary arteries, it will block the blood flow to a substantial part of the heart

This research is a part of the ULTRA-X-TREME project (project number P17-32-ElastiX), which are partly financed by the Netherlands Organisation for Scientific Research (NWO).
Moein Mozaffarzadeh, Anne E.C.M. Saris, Jan Menssen and Chris L. de Korte are with the Medical UltraSound Imaging Center, Department of Medical Imaging, Radboud University Medical Center, 6500 HB Nijmegen, The Netherlands. Chris L. de Korte is also with the Physics of Fluids Group, MIRA, University of Twente, 7500 AE, Enschede, The Netherlands.

email: moein.mozaffarzadeh@radboudumc.nl , Anne.Saris@radboudumc.nl , Jan.Menssen@radboudumc.nl ,chris.dekorte@radboudumc.nl.

muscle and cause a heart attack.

Stable and rupture-prone plaques primarily differ in their geometry and composition. While ultrasound (US) B-mode imaging can be used to visualize the plaque geometry and the presence of calcifications, accurately distinguishing between these two types of plaque is still a great challenge in cardiovascular research [7]. Vascular strain imaging has the potential to address this issue [8, 9]. In lipid-rich plaques, higher strain values were observed than in plaques with predominantly fibrous material.

Ultrasound elastography (strain) imaging is carried out through estimating the displacement between two frames of beamformed RF data and taking the spatial derivative of the displacement. It was initially introduced back in 1991 by Ophir, et al. [10], and since then it has been used for tissue characterization [11-13], and detecting tumors in liver [14] and breast [15]. It can be also used to acquire functional information in vascular applications [7, 16, 17]. In this application, the pulsatile behavior of the blood pressure results in the arterial wall deformation. If a plaque is present, the degree of deformation in this plaque can be quantified with vascular strain imaging which was proven to be correlated to the local plaque composition [17, 18].

After the introduction of the concept of US strain imaging [10], different methods have been published to improve this imaging technique [12, 19-24]. The fundamental mechanical limitations of elastography were investigated in [25, 26]. Global stretching [27] and temporal stretching [28] of the radio frequency (RF) signal along with multi-compression averaging [29] and a deconvolution filter [30] were proposed to decrease the decorrelation artifact and therefore increase the signal-to-noise (SNR) in elastograms (strain images). The Least-SQuares Strain Estimator (LSQSE) was introduced by Kallel and Ophir [31] where it was shown that LSQSE significantly increases the SNR of elastograms compared to taking the gradient of displacement for generating the elastograms. The SNR and accuracy of the elastograms were compared when the cross correlation is applied on the envelope and RF signals [32, 33]. It was shown that for small strains, the RF signals provide a superior estimation. The intricate relationship between the resolution and system parameters such as the kernel and template size of the cross correlation function, transducer frequency, bandwidth and f-number was investigated in [34-36].

For better understanding of the mechanical properties of the plaque, radial and circumferential strains are mainly required, for two reasons: 1) the carotid artery (CA) has a cylindrical shape and the plaque is mostly covering a substantial part of the circumference [37], and 2) the mechanical properties of plaques are not isotropic. The arterial wall is composed of different layers that each have a different build up: the intima is a one-cell-thick-layer while the media is mainly composed of elastin and fibrin strands oriented in different directions. Plaques are also composed of different materials each having their mechanical properties in terms of direction and modulus. Consequently, radial as well as circumferential strain helps to evaluate how the artery wall tissue deforms with the cardiac cycle.

Conventionally, the axial (along the ultrasound beams) and lateral (perpendicular to the ultrasound beams) strain are initially estimated. While the axial displacement/strain can be usually estimated with a high precision, the lateral displacement/strain has a lower precision due to the lack of phase information in the lateral direction of the beamformed RF data. Since both axial and lateral displacements are required to estimate radial and circumferential strain, the image quality and accuracy of these strain components are restricted.

Konofagou and Ophir [38] reported the first high precision estimation of the lateral component of the strain tensor by tracking the phase of the RF A-line in the lateral direction using a weighted interpolation method, provided that the lateral sampling is sufficient. Later, this method was further developed for 3D strain imaging [39]. The axial and lateral displacement (estimated by the method introduced by Konofagou and Ophir [38] ) was used by Fernandez and Xie [40] for estimating the radial and circumferential strain of 12 patients to evaluate the *in vivo* applicability of the technique. The RF signal along multiple beam insonification directions was used to estimate the lateral and axial components of the displacement vector [41]. Later on, the displacement vectors estimated at each angle were weighted based on the insonification angle and spatially compounded to increase the SNR [42]. Both the methods introduced at [41, 42] were based on the tissue incompressibility assumption, which was addressed in [43, 44] by using a least-squares compounding technique instead of the weighted-compounding one. A phased array probe was used in [41-44] and manually rotated to insonify at different directions with a line-by-line imaging sequence. Rao, et al. [45] used the same method while a linear array was used for imaging and beam steering was carried out electronically. Korukonda and Doyley [46] used coherent compounding to improve the point spread function (PSF) in the lateral direction and increase the precision of the lateral displacement.

Our group also developed several techniques to estimate the lateral displacement with a high precision and bring more insight into CA strain imaging. Hansen, et al. [47] proposed to steer the transmit beam at large angles (up to 45°) such that the CA wall could be segmented into circular segments in which the radial strain and the ultrasound beam or the circumferential strain and the ultrasound beam are closely aligned such that they can be estimated by only estimating the axial displacement component. In [48, 49], RF data from a maximum of three beam steering angles were used to estimate the axial and lateral displacement components with a high precision. Adaptive low pass filtering was used to remove the artifacts caused by grating lobes as a consequence of steering the beam at large angles [47, 50]. In [47-49], a line-by-line imaging sequence was used, which limits the data acquisition frame rate, especially in 3D. This issue was addressed in [50, 51] by transmitting plane waves instead of focused waves, at the cost of slight loss of SNR and contrast-to-noise ratio. This method was used for semi-3D strain imaging of a numerical atherosclerotic CA model [52], a bifurcation phantom [53] and four asymptomatic individuals *in vivo* [54], by sweeping a linear array along the elevation direction (longitudinal direction of the CA).

In this paper, we propose an imaging sequence (IS) along with a beamforming strategy to improve the accuracy and robustness of axial and lateral (and thus radial and circumferential) strain estimation by plane waves transmission at different angles and coherently compounding the raw RF data [55] on three sets of beamforming grids. In this way, the phase information is maximized along the axial direction of each of the grids, facilitating our displacement estimation strategy [50, 51]. Through simulations, we have quantified the improvement in displacement and strain estimation accuracy obtained by the proposed IS, originating from the improved full-width-half-maximum (FWHM) and contrast ratio (CR) of the coherently beamformed RF data ([22]), compared to transmitting a single plane wave [50, 51], and finally verified the findings through experiments (CA phantom).

## II. Materials and Method

### A. Displacement estimation strategy

Our displacement estimation strategy is extensively discussed in previous publications [50, 51]. Briefly, it consists of the following steps:

1- Transmission of plane waves at angles of ±20° and 0°, and recording the backscattered wavefronts.
2- Beamforming on grids aligned on ±20° and 0°.
3- Using the RF data beamformed on ±20° to estimating the inter-frame axial displacement in order to compound the inter-frame lateral displacement.
4- Using the RF data beamformed on 0° to estimate the inter-frame axial displacement.

The method presented in [50, 51] is referred as "current" in this paper.

### B. Receive medium

In the original directional beamforming method [56], the grids are defined along the 0°, and the receive sub-aperture determines the beamforming angle. On the other hand, our displacement estimation strategy requires the grids to be defined along an angle (±20° in this paper) to maximize the phase information along the axial direction in the beamformed RF data. To distinguish these two concepts, in this paper, the term "$\theta$° medium" is used to refer to grids aligned along an angle of $\theta$. We used 20°, -20° and 0° mediums (following our displacement estimation strategy [8, 57], explained in Section II.A.); see Figure 1(a). The mediums were defined at ±19.4710° to form mutual intersection points between the angled (±19.4710°) and non-angled (0°) grids [8]. For simplicity, these angles are referred to as ±20° throughout this paper. The center of the mediums had the same coordinate.

### C. Numerical study

Field II ultrasound simulator was used to assemble the simulation data [58, 59]. The GE9LD probe ($f_0$=5.3 MHz, pitch= 230 µm, width= 43.9 mm) was used for the simulations. The pulse repetition frequency (PRF) was 10 kHz. Each set of successive transmissions were simulated at a frame rate of 50 Hz. Two situations were simulated: 1) a single scatterer positioned at the depth of 15 mm and 2) scatterers constructing a cylinder (mimicking the carotid artery (Figure 1(a))). The cylindrical phantom had an inner and outer diameter of 6 mm and 12 mm, respectively, forming a CA wall with a thickness of 3 mm. To mimic human tissue, the rest of the medium (60 mm ×16 mm × 24 mm (lateral × elevation × axial)) was filled with scatterers having an amplitude of 20 dB below that of the CA wall. Ten scatterers per $\lambda^3$ were used to generate a fully developed speckle pattern and thus properly simulate the human tissue and the CA wall; in Figure 1(a), the density of the blue-colored scatterers was deliberately lowered to still be able to recognize the cylindrical phantom. A Tukey window with a cosine fraction of 0.5 was used for apodization of the elements in transmit. To maintain the computation accuracy in Field II, each element of the probe was divided into 2 horizontal and 10 vertical mathematical elements, respectively, and the sampling frequency was set 148.4 MHz. After data acquisition, the raw RF data was downsampled to 21.2 MHz (four times of the central frequency of the probe), in accordance with the sampling frequency of the ultrasound probe in our experimental study.

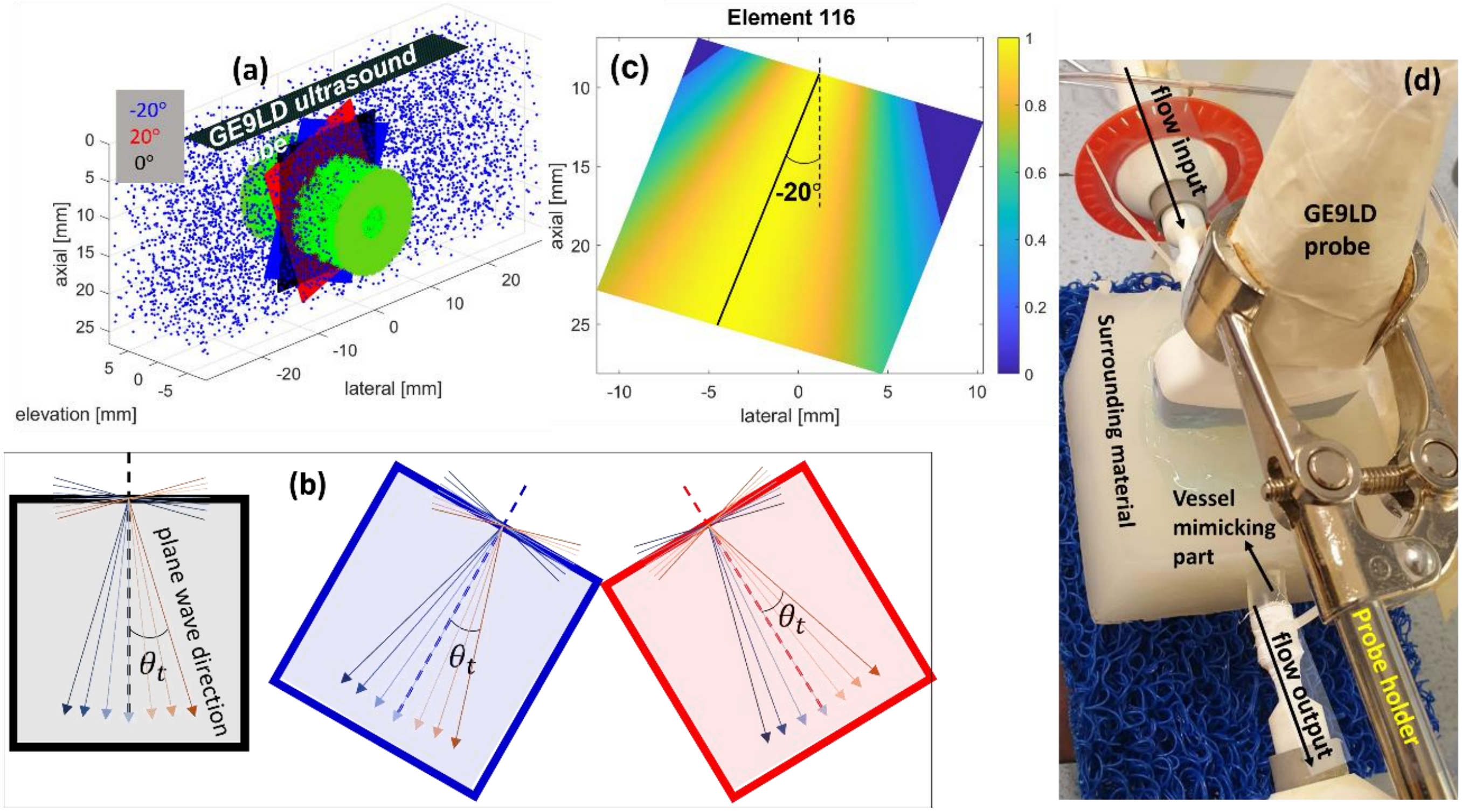

*Figure 1. (a) The numerical imaging setup including the probe, the three angled media, cylindrical phantom in green (mimicking the carotid artery) and the background scatterers in blue (mimicking the tissue surrounding the carotid artery). The density of the blue-colored scatterers was deliberately lowered to be able to recognize the cylindrical phantom. (b) A schematic showing the direction of the plane waves (arrows) with respect to the three sets of media (black, blue and red boxes). The central transmit angle was adjusted with respect to the corresponding grids. (c) The receive grid apodization values of element 116 of the probe on grids defined on 20° where the central line of the element is first corrected (solid line) with respect to the actual central line (dashed line) considering $\theta_t$ = -20° and then the apodization values are assigned. (d) The experimental setup.*

#### 1) Inter-transmit motion

In practice, the carotid artery wall undergoes motion between each transmit event due to the varying blood pressure resulting in a suboptimal coherently compounded signal. The motion between each transmit event is called inter-transmit motion in this paper and can be obtained based on the PRF and the maximum displacement present at the inner diameter of the wall. Physiologically, the area at the lumen vessel-wall boundary of the CA wall will have a displacement of about 10% of the luminal diameter of the wall in a quarter of a second (250 ms) resembling the systolic phase, which results in an inter-transmit motion of 0.12 in radius (0.24 in diameter). This motion was taken into account in the simulations: 1) the single scatterer was moved diagonally between each transmit event, 2) the CA wall was sectioned with a radial step size of 10 µm, and the inter-transmit motion was applied on the location of the scatterers following the 1/R relation between the displacement and the radius of the center of each radial section (R) [9]. The scatterers mimicking the background tissue were static in all the simulations. Only two frames were simulated mimicking the peak systolic phase.

#### 2) Phase information in the axial direction

To compare the phase information in the axial direction, a single scatter was imaged with 2 imaging sequences: 1) single plane wave transmitted at 20° and beamformed on 20° medium, and 2) 5 plane waves transmitted between 10° to 30° and beamformed on 20° medium; The 20° medium was selected because our lateral displacement estimation strategy uses the beamformed RF data on this angle for displacement compounding.

#### 3) PSF evaluation and transmit optimization

To find the desired number of transmit events (i.e., virtual sources (VS) transmitting plane waves), the point scatterer was imaged with ISs comprising different numbers of VS (nVS) within a single sided maximum transmit angle ($\theta_t$) of 5°, 10° and 15°; see Figure 1(b) as an example where nVS was 7. Note that the angles at which the plane wave is transmitted are with respect to the angle of the corresponding receive mediums (Figure 1(b)). Therefore, the backscattered wavefronts corresponding to plane waves transmitted around a $\theta$° medium is only beamformed and coherently compounded on the $\theta$° medium.

The contrast ratio (CR) and full-width-half-maximum (FWHM) along the axial and lateral direction were used for quantitative evaluation of the beamformed RF data. The FWHM is the width of the main lobe of the PSF at -6 dB. The CR was defined as the ratio between the energy in the side-lobes relative to the total energy:

$$CR(r) = 20\log_{10}\sqrt{\frac{E_{out}(r)}{E_{total}}}, \qquad (1)$$

where $E_{total}$ is the total energy in the medium (containing the PSF) and $E_{out}(r)$ is the energy beyond a radius of $r$ which was considered 2.5$\lambda$ [60].

As the pitch of the probe is larger than half of the wavelength, the medium was chosen large enough (16 mm × 26 mm (lateral × axial)) to see the effect of the grating lobes and have it involved in the CR calculation. The absolute CR and FWHM values were reported. The CR of the current and the proposed IS are also evaluated at different SNR levels (from -10 dB to 40 dB) by adding noise to the raw RF data.

### D. Evaluation of the carotid artery simulation

Root-mean-square-error (RMSE) metric, calculated between the estimated displacement/strain and the ground truth, was used for evaluating the results of the CA simulation at different levels of SNR; it was varied from -10 dB to 40 dB by adding noise to the raw RF data.

### E. Experimental study

The experimental setup is shown in Figure 1(d). The phantom was made out of PVA (Poly(vinyl alcohol) and cooling liquid by 40% mixing ethylene glycol (Fluka 03770-5L, technical, suitable as anti-freezing agent) and 60% demi-water. After pouring the mold with the mixture, we waited 2 hours for the remaining air bubbles to escape and to cool the liquid down to room temperature (23 degree), and then started the freeze-thaw cycle. The width of the wall was 2 mm, with an inner diameter of 8 mm. The GE9LD probe was connected to a Verasonics Vantage 256 and positioned perpendicular (cross-sectional view) on the CA phantom (comprising of a vessel mimicking part that is embedded in a surrounding material) with a probe holder. The center of the lumen was positioned at the depth of 15 mm and the phantom was coupled to the probe using ultrasonic gel. A pump (model H3F, Liquifl, Garwood, USA) was used to generate a realistic flow pattern and intraluminal pressure (frequency of 60 bpm, amplitude of 0.6 liter/minute on top of a steady flow of 0.5 liter/minute) resulting in a diastolic and systolic pressure of 70 mmHg to 150 mmHg using a single adjustable flow resistor, respectively. Water was used as the intraluminal fluid. The PRF was 10 kHz with a successive transmissions rate of 50 Hz. Data acquisition was carried out for 3 seconds, resulting in 150 successive frames. Two imaging sequences were used for data acquisition:

1) the current IS: transmitting 3 plane waves at ±20° and 0°; one for each receive medium.

2) the proposed IS: transmitting 57 plane waves with a single sided transmit angle ($\theta_t$) of 10°; 19 plane waves for each receive medium. Selecting nVS of 19 and $\theta_t$ of 10° will be justified in Section III.B.

The axial and lateral displacement estimates along with the max and min principal strains were provided. The first frame acquired at the beginning of the heart cycle was considered as the reference for the inter-frame displacement estimation.

### F. Beamforming

Delay-and-sum was used to beamform the raw radio frequency (RF) data [61, 62]. The intensity seen by each element at each grid (pixel) $Q$ with a coordinate of $R, \theta$ was weighted ($P_Q$) using the directional sensitivity of the element (equation 8.34 of reference [63]):

$$P_Q(R,\theta) \sim sinc\left(k\frac{L}{2}\sin(\theta)\right). \quad (2)$$

Where $k$ is the wavenumber, and $L$ is the element width. For the angled mediums (±20°), beamforming along angled grids is needed to preserve the phase information along the axial direction [56]. This requires correcting the directional sensitivity of each element with respect to the angle of the medium (i.e., the center line perpendicular to each element should be rotated according to the angle of the medium (see Figure 1(c) where the weights calculated for -20° medium and element 116 of the probe is shown)). The field of view of each element was considered 30° and therefore, the weight of the grids located at angles above 30° was 0 [64]. Note that this is an alternative approach of using a dynamic F-number (F#) in the beamformer where the opening angle is related to the F# by $F\# = 1/(2 * \tan(FOV))$, where $FOV$ is the field of view [65].

For beamforming, a grid size of 18 µm and 46 µm was used in the axial and lateral directions, respectively. We selected the grid size such that a circular-shaped cross-correlation function peak is obtained when estimating the displacement using the 2D speckle tracking method [66, 67].

### G. Two-step speckle tracking

Tissue displacement along the axial direction was estimated through a two-step procedure [68, 69]. First, the envelope of the beamformed RF data was used to estimate the global axial displacement. Second, the beamformed RF data was used to find the subsample displacement estimates by interpolating the cross-correlation-function (CCF). The output of the first step was first median filtered and then used as the input of the second step, and finally the output of the second step was median filtered. For the first step, a 2D kernel size of 0.8 mm × 0.8 mm (axial × lateral) with a 2D search margin of 0.2 mm × 0.2 mm and median filter window size of 0.3 mm × 0.3 mm was used. For the second step, a kernel size of 0.2 mm × 0.2 mm (axial × lateral) with a search margin of 0.06 mm × 0.1 mm and median filter window size of 0.3 mm × 0.3 mm was used. The final resolution of the displacement estimates was 0.152 mm × 0.054 mm (105 × 393 grid).

### H. Strain estimation

For estimating the axial and lateral strain, the first-order spatial derivative of the axial and lateral displacement along the axial and lateral directions were calculated by LSQSE (introduced by Kallel and Ophir [31]), respectively. A window size of 1.52 mm × 1.08 mm (10 grid × 20 grid) and 0.76 mm × 1.62 mm (5 grid × 30 grid) was used to estimate the strains from the displacement estimates in the axial and lateral directions using the LSQSE, respectively. The principal components were determined to obtained the radial and circumferential strains (minimum and maximum principal components, respectively) [70].

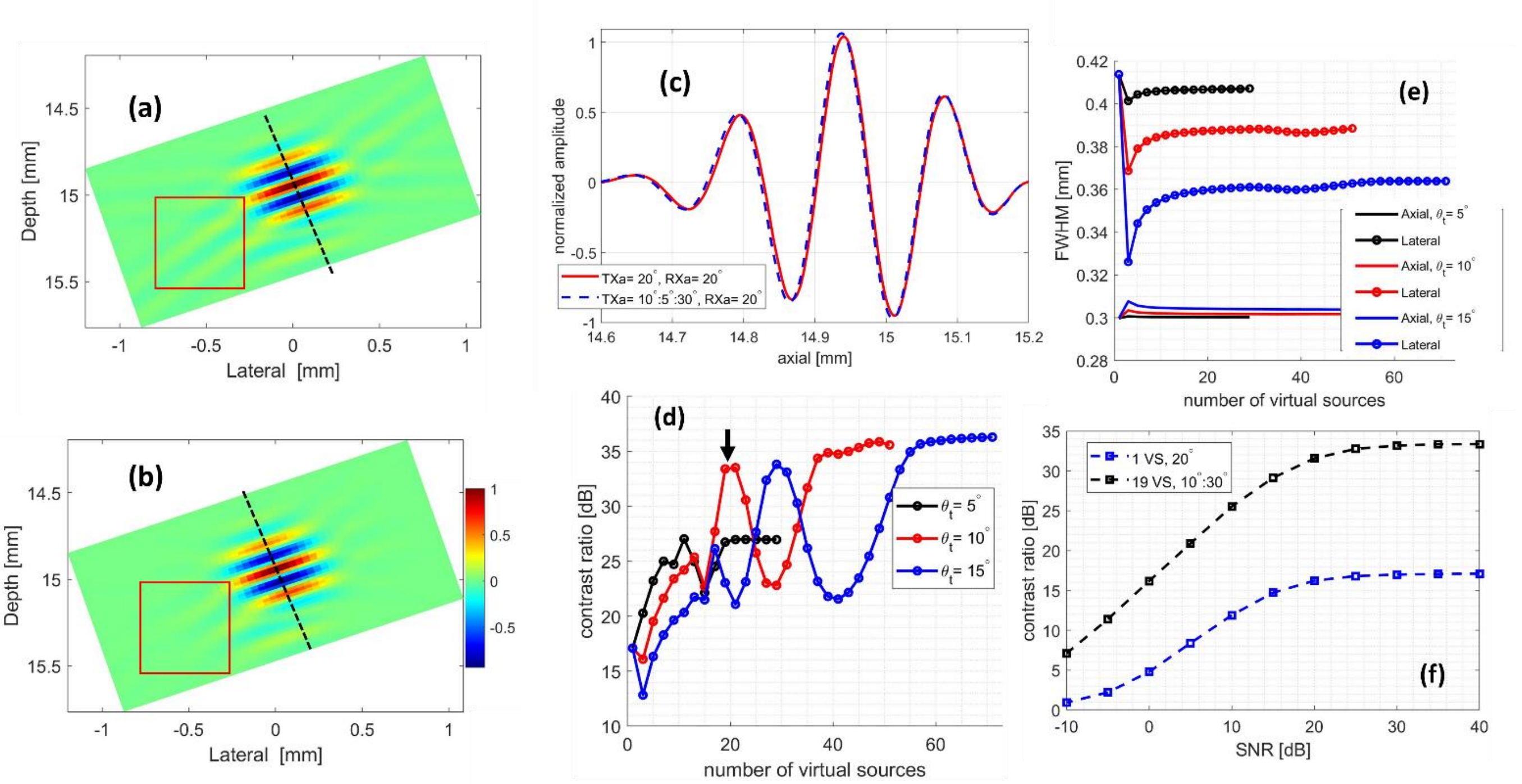

*Figure 2. The point spread function (PSF) when plane wave(s) is(are) transmitted and beamformed at (a) 20° and 20° medium, and (b) 10°:5°:30° and 20° medium, respectively. Beamformed RF data before envelope detection was used for PSF visualization. (c) The comparison between the phase information in the axial direction of the PSFs shown in (a, b) where TXa and RXa indicate the transmit and medium angle, respectively. (d) The contrast ratio (CR) and (e) the lateral and axial full-width-half-maximum (FWHM) for different number of virtual sources and single sided max transmit angle ($\theta_t$) of 5°, 10°, and 15°. In (d, e), CR change was negligible after the maximum number of virtual sources used for each graph and therefore not plotted for better visualization. (f) A comparison between the CR of the proposed (19 virtual sources with a single sided max transmit angle of 10°; 19 VS, 10°: 30°) and current imaging sequences (1 VS, 20°).*

## III. Results

### A. Numerical results

#### 1) Phase information single scatterer

As can be seen in Figure 2(c), the phase information of the PSF beamformed on a 20° medium when 5 plane waves centered at 20° (10°: 5°:30°) are transmitted closely follows the phase information of the PSF obtained when a single plane wave at 20° is transmitted. There is a slight shift between them which could be due to discretization and different frequency content originating from coherent compounding. This effect is identical both before and after tissue displacement, and therefore it does not impose any implication on displacement estimation. Coherent compounding decreases the sidelobe level with respect to single angle transmission (compare the intensity inside the red boxes in Figure 2(a) and Figure 2(b)) [55].

#### 2) PSF evaluation single scatterer

As seen in Figure 2(d), the local maximum of the CR values occurs at nVS of 11, 19 and 29, for $\theta_t$ of 5°, 10° and 15°, respectively; fluctuations of the graphs are due to the artifacts caused by the grating lobes. CR gets noticeably improved using $\theta_t$ of 10° and 15°, compared to 5°, due to the relatively additional constructive/destructive interferences of the mainlobes/sidelobes caused by transmitting at different angles. The CR values become saturated in the end of each graph since the difference between the transmit angles becomes minor, with no additional constructive/destructive interferences of the mainlobes/sidelobes.

Figure 2(e) shows that the axial resolution change is minor between different $\theta_t$ while the lateral resolution improves (FWHM decreases) with a larger $\theta_t$. The lower FWHM at nVS of 3 is due to the constructive interferences of the mainlobes insonifying the scatterer from different angles. The FWHM increment after a nVS of 3 is due to excessive overlap of the mainlobes.

We choose a $\theta_t$ of 10 with an nVS of 19 as our desired transmit sequence because although the CR differs between the first local maximum of the graphs and where it becomes saturated (the best CR), 19 transmissions is a trade-off between data acquisition frame rate, transfer and processing speed and CR improvement. The lateral resolution of an nVS of 19 with a $\theta_t$ of 10° is worse than nVS of 29 with a $\theta_t$ of 15°, but it is not a critical factor considering our lateral displacement strategy (see references [50, 51] and Section II.A) and the sub-pixel interpolation process for finding the peak of the cross correlation function.

Figure 2(f) shows the CR of the proposed IS compared to the current one ([8, 57]) at different levels of SNR. At the SNR of 0 dB, the proposed IS provides a CR of 16.1 dB, which is about 11.4 dB higher than the current one. Even at the higher range of SNR (40 dB, where the CR gets saturated), the proposed IS improves the CR by 16.3 dB, compared to the current IS.

#### 3) Displacement/strain estimation error CA mimicking vessel phantom

Figure 3(a) and Figure 3(b) show the RMSE of the displacement and strain estimates (with respect to the ground truth) at different levels of SNR, respectively. While the RMSE of the current IS in lateral displacement estimation significantly increases at -5 and -10 dB SNR (about 3.3 and 10 times larger RMSE compared to that at 0 dB SNR, respectively), the RMSE of the proposed IS only increases by 11% and 34%, respectively. The proposed IS lowers the RMSE in lateral displacement estimation by 96 %, at the low levels of SNR. The same trend can be observed in the RMSE graph of the strain estimates. At SNR≥ 20 dB, the RMSE of the axial displacement/strain estimates of both the imaging sequences show to be fairly constant. However, the proposed IS improves (decreases) the RMSE of the lateral displacement and strain estimation by 40 % and 45 %, respectively.

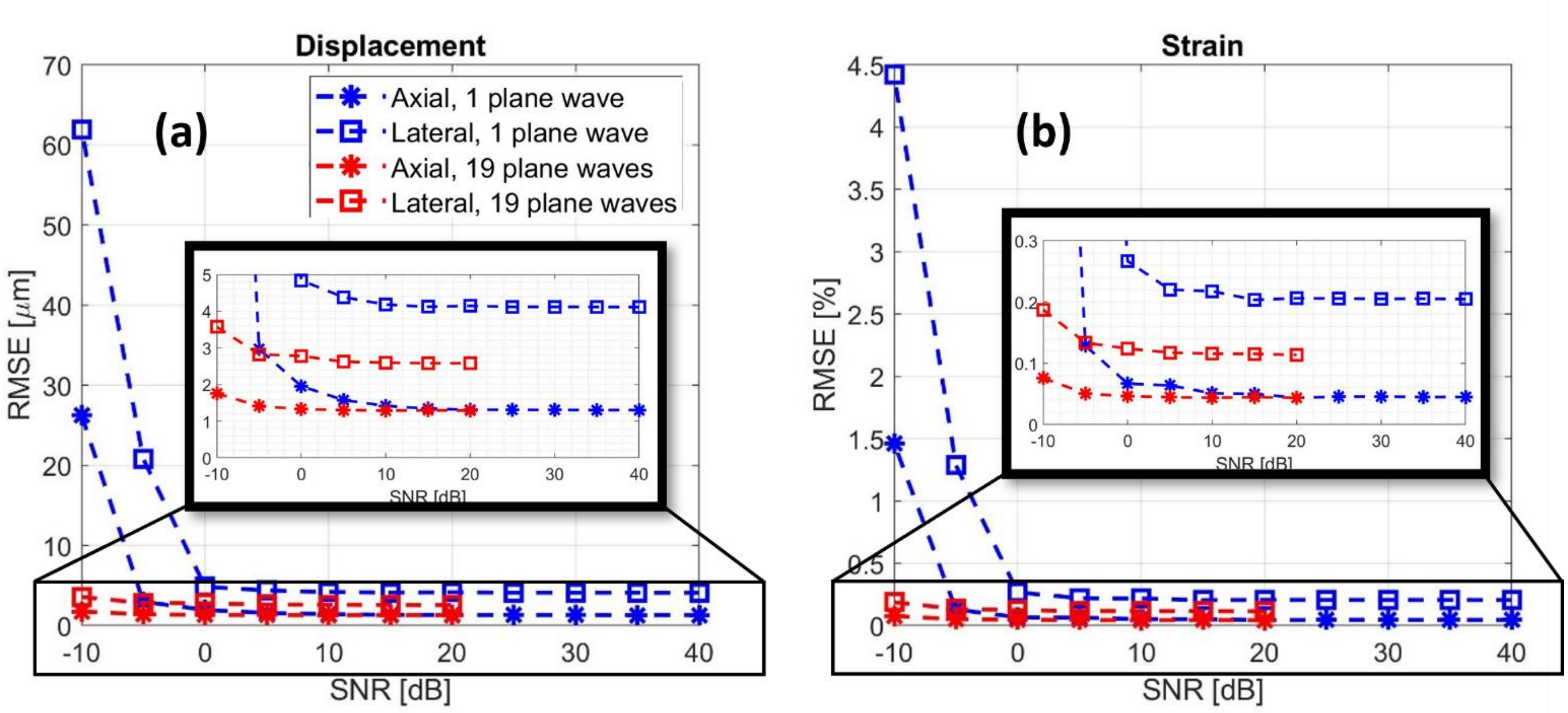

*Figure 3. The root-mean-square-error (RMSE) of the (a) displacement and (b) strain estimations, at different levels of signal to noise ratio (SNR).*

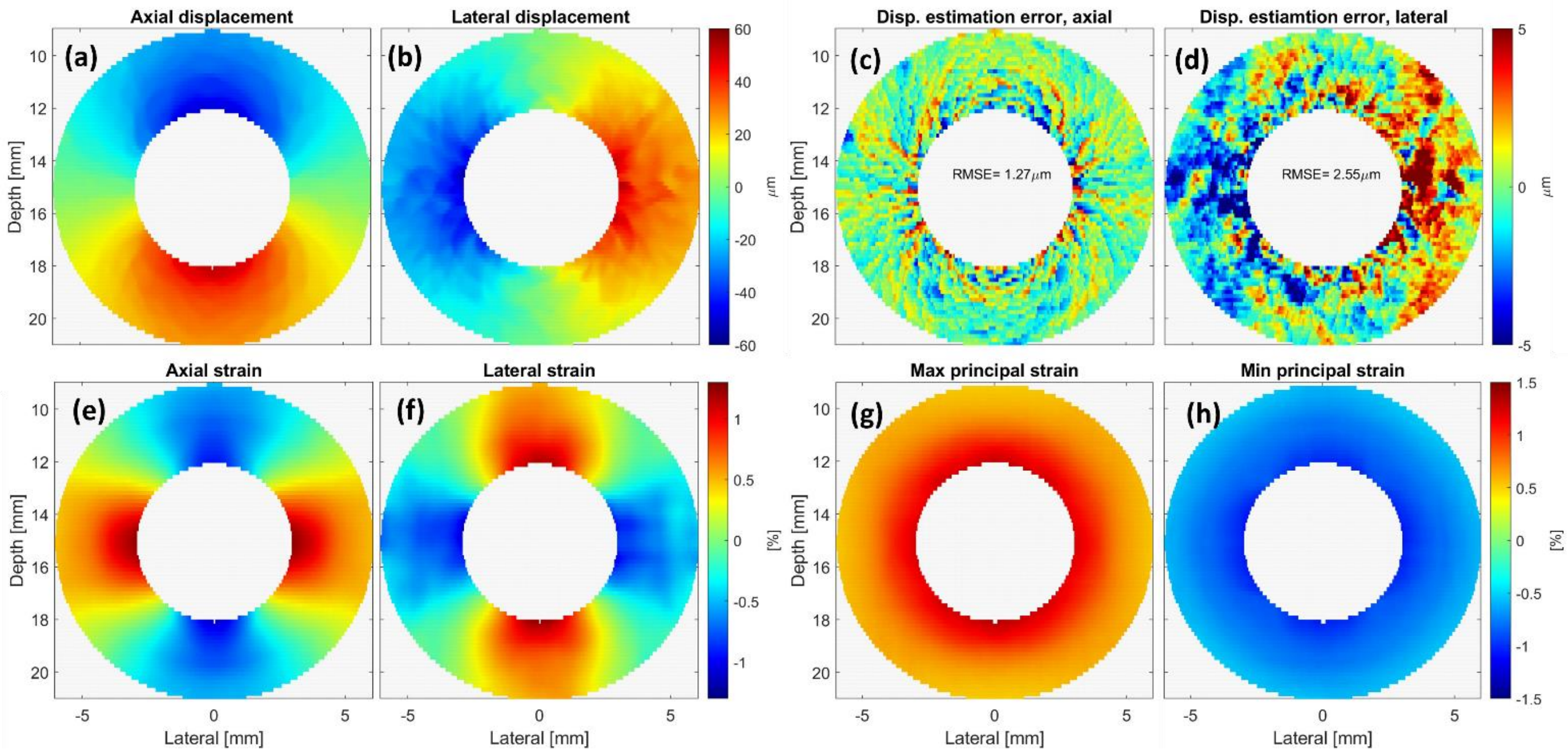

*Figure 4. The estimated (a) axial and (b) lateral displacement, along with the error maps shown in (c) and (d), respectively. The estimated (e) axial and (f) lateral strain. (g) Max principal (circumferential) and (h) min principal (radial) strains. The axial and lateral strains shown in Figure 3(e, f) were used to estimate the min and max principal strains. All the results were obtained using the proposed imaging sequence and a SNR of 20 dB.*

Figure 4(a, b) shows the axial and lateral displacement estimations obtained with the proposed IS when the SNR was 20 dB. The corresponding displacement error maps and the strain images are shown in Figure 4(c, d) and Figure 3(e, f), respectively. The corresponding estimated maximum and minimum principal strains (indicating circumferential and radial strains, respectively) are also provided in Figure 4(g, h).

### B. Experimental results

Slide 1 to 3 of the supplementary document show the B-mode images and displacement/strain estimates over three cardiac cycles. The inter-frame displacement and strain estimations obtained by the current and proposed IS at the systolic phase are provided in Figure 5 and Figure 6, respectively.

The artifact seen at the 5 and 7 o'clock in Figure 5(a) is due to wave diffraction due to the large acoustic impedance mismatch between water and the PVA phantom at the boundary between the inner surface of the lumen and flowing water. This artifact is well addressed by the proposed IS (see 5 and 7 o'clock in Figure 6(a)) because the non-insonified regions in 0° transmission are symmetrically insonified by transmission at other angles centered at 0°. This artifact causes a high strain region at 5 and 7 o'clock of Figure 5(d), which is well suppressed by the proposed IS (see Figure 6(d)).

The proposed IS improves the lateral displacement estimation as can be observed by a less noisy and more gradual transitioning from blue to red compared to the current IS (compare the region between 11 to 1 o'clock and the black box in Figure 5(b) with Figure 6(b)). While the artifact caused by the wave diffraction is barely visible in the lateral displacement obtained by the proposed IS (see 5 and 7 o'clock at Figure 6(b)), there is still a large displacement gradient, which results in the artifacts on 5 and 7 o'clock of the max principal strain shown in Figure 6(c). Yet, this artifact is less prominent than the current IS.

Comparing Figure 5(e) and Figure 6(e) indicates that the proposed IS provides a more accurate strain estimation as it detects the movement of the wall in the secondary phase of the heart pressure cycle. Moreover, the standard deviation of the min principle strain obtained by the proposed IS (0.55%) is less than that obtained by the current IS (0.75%), specially at the peak systolic phase. While the ground truth displacement and strain values are not available, the estimates are as expected and robust through the cycles (see
slide 2 and 3 of the supplementary document).

## IV. Discussion

### 1) Imaging scheme

This paper introduces a new IS to improve strain imaging, by using coherent compounding in combination with displacement compounding [68, 69]. The improvements in the displacement and strain estimations, especially in the lateral direction, are mainly because of the higher level of contrast ratio of the coherently compounded beamformed RF data.

Korukonda and Doyley [46] also used coherent compounding to improve the displacement estimation in the lateral direction by improving the PSF in the lateral direction. However, the lateral displacement was directly estimated by cross-correlating the kernel and template windows on the RF data beamformed on a 0° medium, with no phase information in the lateral direction. Consequently, the second iteration estimation (sub-resolution displacement estimation) in the lateral direction was not carried out with a high accuracy. Our evaluation indicates that the RMSE of the lateral displacement as estimated by our displacement compounding strategy is 2.55 µm (see Figure 3(f)), which is 70% lower than that resulted by coherently compounding of the 19 plane waves on 0° medium (results are not presented in this paper).

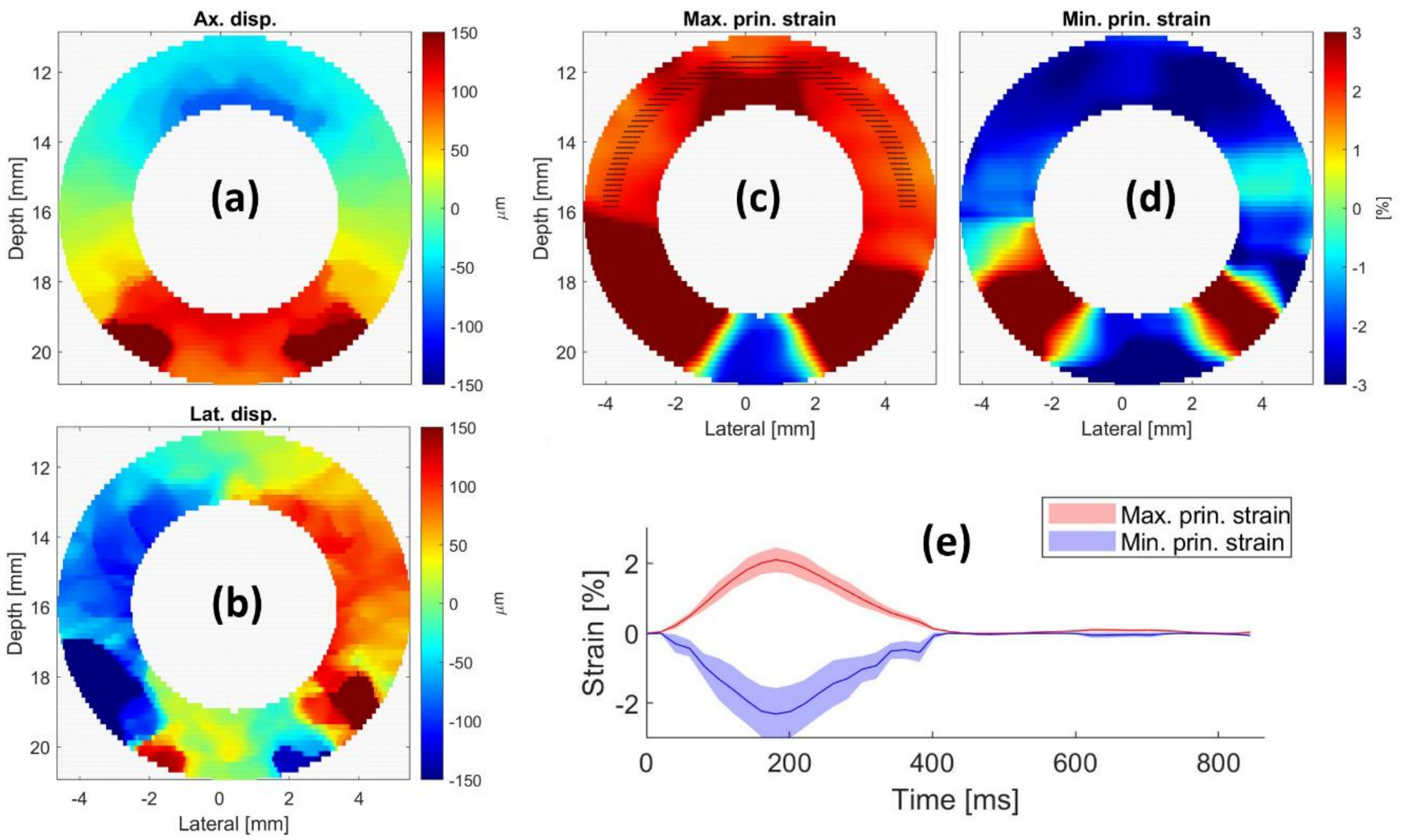

*Figure 5. The experimental results obtained by the current imaging sequence. (a) axial and (b) lateral displacement estimation. The mean (line graph) and the standard deviation (shaded area around the line graph) of the (c) max and (d) min principal strains were measured on the dot-filled area shown in (c), and then shown in (e).*

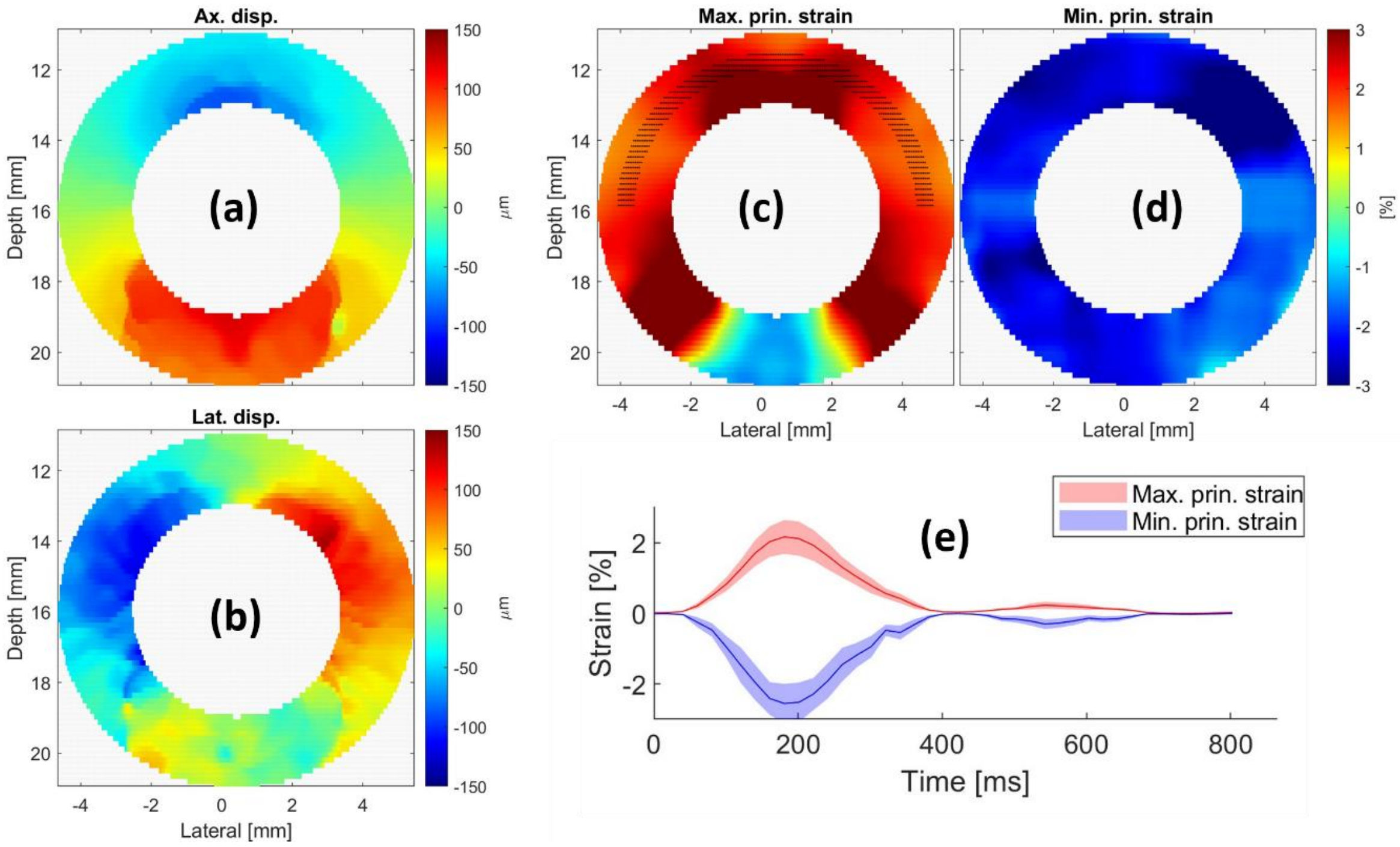

*Figure 6. The experimental results obtained by the proposed imaging sequence. (a) axial and (b) lateral displacement estimation. The mean (line graph) and the standard deviation (shaded area around the line graph) of the (c) max and (d) min principal strains were measured on the dot-filled area shown in (c), and then shown in (e).*

The main limitations of the method proposed by Konofagou and Ophir [38] are the extensive lateral interpolation between the RF lines, the iterative nature of the algorithm itself, line-by-line imaging scheme (which comes with a low data acquisition frame rate) and dependency of the lateral displacement estimation accuracy on the lateral-resolution [71]. Our displacement estimation strategy uses high frame rate plane wave transmission and the axial phase information for estimating the axial displacement on the angled mediums with subsequent tri-angulation to derive the lateral displacement. It mainly consists of beamforming the backscattered RF data on the (non-)angled mediums and cross correlating the kernel and template windows, which both can be implemented real-time on graphical processing units (GPU) [72, 73].

We compared the proposed IS with the second local maximum seen in Figure 2(d) (nVS= 29 and a $\theta_t$=15°) through a simulation study. The RMSE of the axial displacement estimation was the same as that of proposed IS, but with 0.17 µm higher RMSE in estimating the lateral displacement. This could be due to the higher artifacts caused by the grating lobes, compared to 19 VSs, and its additional effect on compounding the lateral displacement using the axial displacement estimations on ±20° mediums.

We have included the effects of grating lobes in our PSF evaluation in order to find the ideal number of virtual sources (nVS) and max one-sided transmit angle ($\theta_t$) as plane wave transmission is carried out using large angles. Suppressing them by our filtering technique (developed by Hansen, et al. [74]) compresses the graphs seen in Figure 2(d) with lowering the local maximum by 10 dB and minimum by 1 dB, but still leads us to a nVS of 19 and $\theta_t$=10°. Note that the PSF evaluation and the proposed transmit settings are limited to the GE9LD probe. The same procedure can be used to find the optimal transmit settings for other probes.

We have based our selection procedure primarily on CR (and not lateral FWHM) for to two reasons: first, the effect on CR of the beamformed RF data is substantial and has more impact on strain compared to minor changes in the FWHM [22], and second, our lateral displacement estimation strategy does not significantly depend on the lateral resolution (such as the method proposed by Konofagou and Ophir [38]).

### 2) Data acquisition frame rate

The data acquisition frame rate of the current IS is 10 kHz, which is 19 times higher than the proposed IS. However, such a high frame rate is not needed for CA strain imaging considering the ranger of the motion of the CA wall.

Compared to a line-by-line IS, the proposed IS provides 17.5 times higher data acquisition frame rate. Using 3 transmit focus and a half-lambda spacing for each focused line (resulting in 108 lines for a 16 mm lateral field-of-view), the PRF reduces to 30 Hz (10,000/(108×3)) for acquiring the raw RF data corresponding to a single medium, while that is 526 Hz (10,000/19) for the proposed IS. The difference between the frame rates will be even larger in 3D imaging as the transmit focal points of the line-by-line sequence should be distributed in a 3D medium.

In case the raw RF data is averaged 19 times by transmitting 19 plane waves on the same transmit angle ($\theta_t$= 0°), the data acquisition frame rate will be the same as the proposed IS. The averaging increases the SNR by 13.8 dB, but the PSF is further improved by the proposed IS due to the constructive/destructive interferences of the mainlobes/sidelobes caused by transmitting at different angles. The impact of the interferences becomes larger at the low SNR levels.

### 3) Inter-frame motion

While we have incorporated the inter-frame motion (resembling vessel wall motion during the systolic phase) in our simulation study, the RF data beamformed on -20° with and without the inter-transmit motion involved in numerically imaging a single scatterer was quite similar; a CR and lateral FWHM difference of 0.03% and 0.01% was observed, respectively, with an equal axial FWHM.

### 4) Translation to 3D

This study was limited to 2D imaging where the elevation focus of the 1D probe affects the quality of the beamformed RF data and thus displacement/strain estimates. 3D quasi-static elastography was recently reported by Papadacci, et al. [75] where a matrix transducer was used for data acquisition. However, only axial strain were reported. Fekkes, et al. [8] compared the performance of the 2D versus 3D displacement estimation in terms of radial and circumferential strain through a series of simulated US images of a patient-specific 3D atherosclerotic CA model and experimentally obtained series of 2D cross-sections in volunteers. It was concluded that the 3D search kernel is needed for accurate estimation of the displacement, especially in phases with a high inter frame longitudinal motion.

Jensen, et al. [76] have shown the feasibility of 3D flow imaging with a row-column array [77, 78] where 12-16 virtual sources were used to get a decent image quality in the plane perpendicular (parallel) to the transmitting (receiving) elements. While it is not possible to simply adapt the current IS (introduced by our group [68, 69]) for 3D using a row-column array due to using a single virtual source for data acquisition and beamforming on each of the 3 mediums, the proposed IS uses multiple virtual sources for each medium and thus might provide sufficient back-scattered raw RF data to get a decent image quality in both the lateral and elevational directions (with respect to the surface of the row-column array); see the slide 4 of the supplementary document where the proposed IS can used to improve the image quality in the elevational direction (perpendicular to the direction of the transmit elements). In our future studies, we will translate the proposed IS in this paper to 3D and use a row-column array to make high quality 3D strain imaging feasible.

### 5) Beamforming

The effect of coherence-based beamforming on the axial and lateral displacement estimation was reported to be negligible [79], and Hendriks, et al. [80] reported that Lu's-fk beamforming results in the most accurate displacement estimation. However, we have used delay-and-sum for beamforming due to its simplicity and ease of translation to 3D when a matrix [64, 81] or row-column array [77, 78] is used for data acquisition. Efficient implementation of delay-and-sum on GPU is also more straightforward than frequency domain beamformers.

### 6) Phase aberration

Phase aberration effects in elastography were investigated in [40, 82]. Varghese, et al. [82] concluded that if the phase aberration is present in both the pre and post compression signals, the precision of the strain estimator is not reduced. However, this conclusion only applies to imaging scenarios including weak aberrating layers, such as carotid artery imaging where skin, CA wall and blood have different wave-speed and density and induce a slight amount of phase aberration. Recent studies have shown that the PSF gets significantly distorted at the presence of a strong aberrating layer (such as human temporal bone [62, 73]), and thus the speckle pattern changes [83]. While the change in speckle pattern still applies to both the pre and post compression frames, the overall structure of the artery under investigation becomes distorted. Moreover, speckle pattern decorrelation occurs as the phase aberration depends on spatial location. These two factors degrade the performance of the cross-correlation function and thus the displacement/strain estimates. As one of our future studies, we will combine the IS proposed in this paper with sophisticated aberration correction techniques [62, 73, 83] to make transcranial strain imaging feasible. This system could be very helpful for determining the rupture risk of intracranial aneurysms [84, 85].

## V. Conclusion

In this paper, we introduced an imaging sequence to improve non-invasive 2D strain imaging. It comprised of transmission of plane waves and coherent compounding of the raw RF data on 3 receive mediums (with grids aligned in 0° and ±20°) required to compound the lateral displacement out of the axial displacement estimated on the angled mediums. Using the numerical data with a SNR≥ 20 dB, the proposed imaging sequence improved the contrast ratio by 16.3 dB and reduced the root-mean-square-error of the lateral displacement and strain estimations by 40% and 45%, compared to the current one, respectively. The experimental results obtained by the proposed imaging sequence were in agreement with the theory and illustrated higher accuracy in strain estimation.

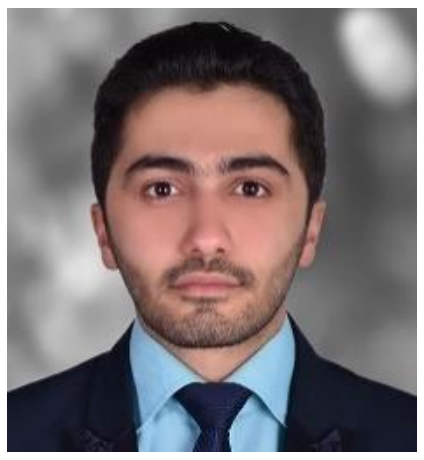

**Moein Mozaffarzadeh** received the BSc degree in electrical engineering from Babol Noshirvani University of Technology, Iran, in 2015, the MSc degree in biomedical-bioelectric engineering from Tarbiat Modares University, Iran, in 2017, and the PhD degree in the department of applied sciences, laboratory of Medical Imaging (in the field of medical ultrasound imaging) from Technical University of Delft, The Netherlands, in 2022. He is currently a PostDoctoral researcher at Medical UltraSound Imaging Centre at the Department of Medical Imaging: Radiology of Radboud university medical center. His research interests include transcranial photoacoustic/ultrasound imaging, acoustic beamforming and medical ultrasound system development.

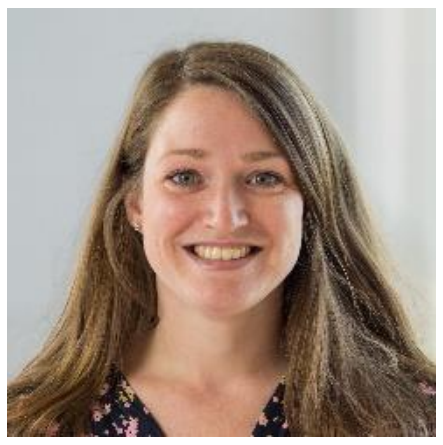

**Anne E.C.M. Saris** received her M. Sc. degree in Medical Engineering (2012) at the Eindhoven University of Technology, The Netherlands. She did her graduation project at the Radboud university medical center, Nijmegen, The Netherlands on correlation-based discrimination between myocardial tissue and blood in 3D echocardiographic images, for segmentation purposes. Since 2012, Anne has been employed by the Medical UltraSound Imaging Center of the Radboud University Medical Center, Nijmegen, the Netherlands. In 2019, she obtained her Ph.D. degree on blood velocity vector imaging in the carotid artery using ultrasound. Since then, she has been employed as post-doctoral researcher. Her research interests include ultrafast ultrasound, 3D ultrasound imaging, technical development and clinical implementation of ultrasound blood flow imaging, with the overall aim to improve diagnosis, intervention and monitoring of cardiovascular diseases.

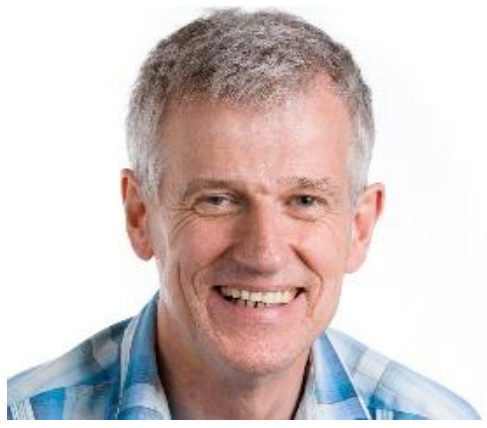

**Jan Menssen** received his Bachelor in in Electric Engineering in 1984. After graduating, he started his career at the Radboud University Medical Center in Nijmegen, the Netherlands as a research assistant at the Obstetrics department. Since then he has been employed at different departments at this medical center. Since 2005 he is a member of the Medical UltraSound Image Center (MUSIC). His research interests were fetal ECG, fetal pulse-oximetry, Surface EMG, NIRS and Ultrasound and he has co-authored more than 10 peer reviewed articles in these fields. At the MUSIC department he assist researchers in their project. In 2022 he received a MSc in Human Movement Science at the Vrije Universiteit Amsterdam.

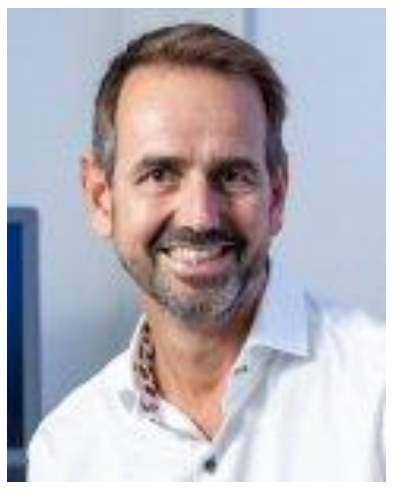

**Chris L. de Korte** (Fellow IEEE) received the M.Sc. degree in electrical engineering from the Eindhoven University of Technology, Eindhoven, The Netherlands, in 1993, and the Ph.D. degree in medical sciences from the Thoraxcenter, Erasmus University Rotterdam, Rotterdam, The Netherlands, in 1999, with a focus on intravascular ultrasound elastography. In 2002, he joined the Clinical Physics Laboratory, Radboud University Medical Center, Nijmegen, The Netherlands, as an Assistant Professor and became the Head in 2004. In 2006, he was appointed as Associate Professor of medical ultrasound techniques and he finished his training as a Medical Physicist in 2007. Since 2012, he has been the Founder and the Director of the Medical UltraSound Imaging Center (MUSIC), Department of Medical Imaging, Radboud University Medical Center (Radboudumc), Nijmegen, The Netherlands. He was appointed as a Full Professor on medical ultrasound techniques in 2015. Since 2016, he has been a Full Professor of medical ultrasound imaging with the University of Twente, Enschede, The Netherlands. He has (co)authored over 200 peer-reviewed articles in international journals and is (co)inventor of four patents. His research interests include functional imaging and acoustical tissue characterization for diagnosis, treatment monitoring, and guiding interventions for oncology and vascular applications. Recently, he also focuses on artificial intelligence-driven apps for point-of-care ultrasound. Dr. de Korte is recipient of the EUROSON Young Investigator Award 1998 of the European Federation of Societies for Ultrasound in Medicine and Biology.